\documentclass[11pt,a4paper]{article}
\usepackage{jcappub}
\usepackage{ifthen}
\usepackage{epsfig}

\newcommand{\dd}{\textrm{d}}

\newcommand{\beqra}{\begin{eqnarray}}
\newcommand{\eeqra}{\end{eqnarray}}
\newcommand{\beq}{\begin{equation}}
\newcommand{\eeq}{\end{equation}}
\newcommand{\n}{\ensuremath{\mathbf{n}}}

\title{Non-Gaussianity and CMB aberration and Doppler}
\author[a]{Riccardo Catena}
\author[b,c]{Michele Liguori}
\author[d]{Alessio Notari}
\author[b,c,e]{and Alessandro Renzi}
\affiliation[a]{Institut f\"ur Theoretische Physik, Friedrich-Hund-Platz 1, 37077 G\"ottingen, Germany} 
\affiliation[b]{INFN, Sezione di Padova, Via Marzolo 8, 35131 Padova, Italy}
\affiliation[c]{Dipartimento di Fisica e Astronomia ``G. Galilei",
Universit\'a degli Studi di Padova , Via Marzolo 8, 35131 Padova, Italy}
\affiliation[d]{Departament de F\'isica Fondamental i Institut de Ci\'encies del Cosmos,
Universitat de Barcelona, Mart\'i i Franqu\'es 1, 08028 Barcelona, Spain}
\affiliation[e]{SISSA, Via Bonomea 265, Trieste, I-34136, Italy}

\emailAdd{riccardo.catena@theorie.physik.uni-goettingen.de}
\emailAdd{notari@ffn.ub.es}
\emailAdd{michele.liguori@pd.infn.it}
\emailAdd{arenzi@pd.infn.it}
\abstract{The peculiar motion of an observer with respect to the CMB rest frame induces a deflection in the arrival direction of the observed photons (also known as CMB aberration) and a Doppler shift in the measured photon frequencies. As a consequence, aberration and Doppler effects induce non trivial correlations between the harmonic coefficients of the observed CMB temperature maps. In this paper we investigate whether these correlations generate a bias on Non-Gaussianity estimators $f_{NL}$. We perform this analysis simulating a large number of temperature maps with Planck-like resolution (lmax $= 2000$) as different realizations of the same cosmological fiducial model (WMAP7yr). We then add to these maps aberration and Doppler effects employing a modified version of the HEALPix code. We finally evaluate a generalization of the Komatsu, Spergel and Wandelt Non-Gaussianity estimator for all the simulated maps, both when peculiar velocity effects have been considered and when these phenomena have been neglected. Using the value $v/c=1.23 \times 10^{-3}$ for our peculiar velocity, we found that the aberration/Doppler induced Non-Gaussian signal is at most of about half of the cosmic variance $\sigma$ for $f_{NL}$ both in a full-sky and in a cut-sky experimental configuration, for local, equilateral and orthogonal estimators. We conclude therefore that when estimating $f_{NL}$ it is safe to ignore aberration and Doppler effects {\it if} the primordial map is already Gaussian. More work is necessary however to assess whether a map which contains Non-Gaussianity can be significantly distorted by a peculiar velocity.}


\begin{document}
\maketitle

\section{Introduction}
The issue of the impact of the peculiar motion of an observer on the Cosmic Microwave Background (CMB) is less trivial than it may seem. Its main effect is to induce a large dipole, even on a homogeneous map, and actually the value of the dipole is used to measure our velocity compared to the CMB rest frame. However the Doppler effect and  aberration have an effect on all multipoles $\ell$ when applying a boost to a non-homogeneous map.  
In earlier works \cite{Challinor:2002zh} and more recently \cite{Kosowsky:2010jm, Amendola:2010ty, Notari:2011sb, Pereira:2010dn, Chluba:2011zh} this effect has been studied, in order to assess its magnitude on two-point correlation functions. In particular in \cite{Kosowsky:2010jm, Amendola:2010ty, Notari:2011sb} it has been pointed out that the non-diagonal two-point functions can be used as an alternative method to measure our velocity, while in \cite{Catena:2012hq} it has been found that the diagonal two-point function ({\it i.e.} the power spectrum) can be affected leading to a bias that can go up to two standard deviations on some cosmological parameters, for a Planck-like resolution experiment.

It is natural to wonder also if higher-order correlation functions can be significantly affected. As stressed already in \cite{Challinor:2002zh} the aberration and Doppler transformation is a linear mixing between multipoles and therefore if the initial spectrum is Gaussian then Gaussianity is preserved. 
However there could still be a sizeable shift on a single random realization of a map, especially at large $\ell$, since the effect is a mixing of order $\beta \ell$. More precisely, a Non-Gaussianity estimator is a sum over a huge number of combinations of triangles: for each triangle there is a large aberration/Doppler induced shift in the measured value of its contribution to the total sum, so it is important to check if when summing over many triangles the effect becomes negligible for a given experimental sensitivity and resolution. Moreover, also the variance of the estimators will be affected by the considered mixing matrix. 

In the present paper we use a direct approach: we simulate 120 maps with Planck-like resolution and add to these maps aberration and Doppler effects. We then evaluate for each map estimators for the three-point function, both before including aberration and Doppler effects and after having accounted for these effects, in order to check how much the corresponding Non-Gaussianity parameters can be biased by applying the aberration/Doppler transformation on the simulated maps. In particular we focus on the usual $f_{NL}$ parameters for local, equilateral and orthogonal Non-Gaussianity. We perform our checks both on full-sky maps and on cut-sky experimental configurations.

The paper is organized as follows. In section \ref{ab} we shortly review the aberration and Doppler effects on a CMB map and on CMB correlation functions. In section \ref{data} we describe our procedure to construct simulated maps with and without aberration and Doppler effects. In section \ref{analysis}  we define our estimators and show the results. Finally in \ref{conclusion} we draw our conclusions. 

\section{CMB aberration and Doppler}
\label{ab}
The CMB aberration consists in an apparent deflection of the observed CMB photons induced by the motion of the observer ({\it e.g}. the Planck satellite) relative to the CMB rest frame. The difference between the direction $\hat{\n}'$ assigned to the incoming photons by the moving observer and the analogous direction $\hat{\n}$ measured in the CMB rest frame is called aberration angle ${\bf \alpha}$ and is give by the relation
\begin{equation}
{\bf \alpha} \cdot \hat{\mathbf{z}} \;= \frac{\beta \sin^2\theta }{1+\beta  \cos\theta} \,
\label{aberration}
\end{equation}
which follows directly from the velocity transformation relating the two frames, assuming that the observer moves along the direction $\hat{\mathbf{z}}$ forming an angle $\theta$ with the unit vector $\hat{\n}$. The relative motion between the observer and the CMB also induces a shift in the observed frequencies, {\it i.e.} the usual Doppler effect.

 Aberration and Doppler induced distortions of a temperature map are quantitatively described by the following transformation law 
\begin{equation}
T^{\prime}(\hat{\mathbf{n}}') =  \gamma(1+\beta\hat{\mathbf{n}}\cdot\hat{\mathbf{z}}) T(\hat{\mathbf{n}}) \,
\label{boost}
\end{equation}
which relates the temperature $T^{\prime}$ observed in the moving frame to the temperature $T$ measured in the CMB rest frame. In Eq.~(\ref{boost}) the pre-factor $\gamma(1+\beta\hat{\mathbf{n}}\cdot\hat{\mathbf{z}})$ accounts for the Doppler shift of the photon frequencies and, as a consequence of aberration, the maps $T'$ and $T$ are evaluated at different directions, namely $\n'$ and $\n$ which are related by Eq.~(\ref{aberration}). Expanding in spherical harmonics both 
\beq
T^{'}(\hat{\n}') = \sum_{lm} a'_{lm} Y_{lm}(\hat{\n}')
\eeq 
and 
\beq
T(\hat{\n})= \sum_{lm} a_{lm} Y_{lm}(\hat{\n})
\label{CMBrest}
\eeq 
Eq.~(\ref{boost}) can be rewritten as follows
\begin{equation}
  a'_{\ell^{\prime}m^{\prime}} = \sum_{\ell} \int d\hat{\n}\,a_{\ell m'}
\left[\gamma(1+\beta\hat{\n}\cdot\hat{\mathbf{z}})\right]^{-2}Y_{\ell^{\prime}m^{\prime}}^{*}(\hat{\n}^{\prime})Y_{\ell 
m^{\prime}}(\hat{\n})\,.
\label{boost2}
\end{equation}
This expression can be then used to compute the desired correlation functions, as we will see in the next subsections.

\subsection{Two-point functions}

Eq.~(\ref{boost2}) can be conveniently rewritten introducing an aberration kernel defined by
\begin{equation}     
 K_{\ell'\, \ell\, m} = \int_{-1}^{1}  \frac{\dd x}{\gamma\, (1-\beta x)} \,\tilde{P}_{\ell'}^m(x) \, \tilde{P}_{\ell}^m \! \left(\frac{x - \beta}{1 - \beta  x}\right) \,,
 \label{abkernel}
\end{equation}
where
\begin{equation}
  \tilde{P}_{\ell}^m(x) \;\equiv\; \sqrt{\frac{2\ell+1}{2} \frac{ (\ell-m)!}{(\ell+m)!}} \,P_{\ell}^m(x) \,,
\end{equation}
and $P_{\ell}^m(x)$ are the associated Legendre polynomials. The harmonic coefficients in Eq.~(\ref{order2}) are hence given by
\begin{equation}
  a'_{\ell m}\; =\; \sum_{\ell'} K_{\ell' \, \ell\, m} \, a_{\ell' m}\,.
\end{equation}
The aberration kernel can be used to write in a compact form all the desired correlation functions. $K_{\ell' \, \ell\, m}$ can be estimated: in perturbation theory as explained above, by using fitting formulas based on an approximate solution of the integral ($\ref{boost2}$) as in Ref.~\cite{Notari:2011sb}, by means of the recurrence formula of Ref.\cite{Chluba:2011zh} or, finally, within the numerical approach used in Ref.~\cite{Catena:2012hq} in the context of cosmological parameter estimation. The aberration kernel allows to write the two-point correlation function, namely the covariance matrix, as in the following expression
 \beq
 \langle a'_{\ell_1 m}a'^{*}_{\ell_2 m}\rangle =  \sum_{\ell'_1} \sum_{\ell'_2} K_{\ell'_1 \, \ell_1\, m} K^{*}_{\ell'_2 \, \ell_2\, m} \, \langle a_{\ell'_1 m} a^*_{\ell'_2 m} \rangle\,. 
 \eeq

The integrals in Eq.~(\ref{abkernel}) are numerically challenging to compute as a consequence of the highly oscillatory character of the integrand which is proportional to the product of two spherical harmonics evaluated at different directions. The harmonic coefficients (\ref{boost2}) can be however conveniently estimated by expanding the integrand around $\beta$ equal to zero, being our peculiar velocity small in natural units {\it i.e.} $\beta \simeq 1.23\times 10^{-3}$.  At second order in $\beta$ the relation between the primordial $a_{\ell m}$ in the CMB rest frame and the $a'_{\ell m}$ measured by the moving observer takes the following form 
\begin{equation}
  a'_{\ell\, m} \simeq (1+d_{\ell m})a_{\ell m} + c_{\ell m}^{-}a_{\ell-1\, m}+c_{\ell m}^{+}a_{\ell+1\, m} + d_{\ell m}^{-}a_{\ell-2\, m}+d_{\ell m}^{+}a_{\ell+2\, m}\,,
  \label{order2}
\end{equation}
with coefficients $c_{\ell m}^{+}$, $c_{\ell m}^{-}$,  $d_{\ell m}$, $d_{\ell m}^{+}$, $d_{\ell m}^{-}$ given by
\begin{eqnarray}
  c_{\ell m}^{+} & = & \beta(\ell+1)\, G_{\ell+1,m} \,\nonumber \\
  c_{\ell m}^{-} & = & -\beta\ell\, G_{\ell,m} \nonumber\\
  d_{\ell m} & = & \frac{\beta^2}{2} \left[ (\ell+1) (\ell+2) \, G_{\ell+1, m}^2 + \ell (\ell-1) \, G_{\ell, m}^2  - \ell (\ell+1) + m^2-1\right] \nonumber \\
   d_{\ell m}^{+} & = & \frac{\beta^2}{2} \left[ (\ell+1) \,G_{\ell+1, m} (\ell+2) \,G_{\ell+2, m} \right] \nonumber \\
   d_{\ell m}^{-} & = &  \frac{\beta^2}{2} \left[ \ell \,G_{\ell, m} (\ell+1) \,G_{\ell+1, m} \right] 
\label{Ccoef}
\end{eqnarray}
and the function $G_{\ell,m}$ defined as follows
\begin{equation}
G_{\ell,m} = \sqrt{\frac{\ell^{2}-m^{2}}{4\ell^{2}-1}} \,.
\end{equation}
We refer the reader to Refs. \cite{Challinor:2002zh,Kosowsky:2010jm,Amendola:2010ty,Notari:2011sb} for an explicit derivation of such expressions. From Eq.~(\ref{order2}) one finds for the angular power spectrum measured by the moving observer $C^{\prime}_{\ell} \equiv  \langle a^{\prime}_{\ell m}  a^{\prime *}_{\ell m}  \rangle$ the following expression
\begin{eqnarray}
C^{\prime}_{\ell} &=& \sum_{\ell'} C_{\ell'}\Bigg\{ \delta_{\ell \ell'} \left(1- \frac{1}{3}\beta^2(\ell^2+\ell+1)\right)
+ \delta_{\ell(\ell'+1)}\beta^2 \frac{\ell^3}{3(2\ell+1)} \nonumber\\
&+& \delta_{\ell(\ell'-1)} \beta^2\frac{(\ell+1)^3}{3(2\ell+1)} \Bigg\}\,.
\label{Cl}
\end{eqnarray}
Thus, the leading corrections to the primordial angular power spectrum $C_{\ell} \equiv  \langle a_{\ell m}  a^{*}_{\ell m}  \rangle$ are small, {\it i.e.} $\mathcal{O}(\beta^2 l^2)$ (for $l\ll1/\beta$), and tend to zero in the limit of flat angular power spectrum, when $C_\ell \sim C_{\ell+1}$.  This is the reason why in the literature aberration has been neglected in the cosmological parameter estimation: it only mildly affects the observed $C^{\prime}_l$ given in Eq.~(\ref{Cl}). Eq.~(\ref{Cl}) is however correct only in the full sky approximation, which does not account for any foreground contamination of the observed maps. In a more realistic computation, in fact, one has to consider that only a fraction of the sky is actually experimentally accessible. This can be done by multiplying the original full sky map by an opportune window function to mask the background dominated portions of the sky. This procedure leads to a fit of the data based on the concept of pseudo-$C_l$, which, as shown in Ref.~\cite{Catena:2012hq}, are significantly more affected by CMB aberration and Doppler. Moreover, for  $\ell\sim1/\beta$ the perturbative approach is no longer correct. This implies that at large $\ell$ a more sophisticated description of aberration and Doppler effects is required in order to estimate the impact of these phenomena on observable quantities, in particular since the Planck satellite very soon will be able to probe the regime $\ell\sim1/\beta$. In Ref.~\cite{Catena:2012hq}, following a numerical approach similar to the one used in the present paper (and explained below), it has been shown that aberration and Doppler can actually significantly affect a CMB based cosmological parameter estimation. In that work the impact of neglecting Doppler and aberration has been estimated in a Likelihood based analysis of temperature maps simulated from a fiducial model including Doppler and aberration. In a few simulated maps it was possible to show that neglecting these effects induces a bias on the cosmological parameters as large as almost two standard deviations.

 Aberration and Doppler do not only affect the angular power spectrum (see Eq.~(\ref{Cl})), but also the off-diagonal correlation functions, namely the terms  $\langle a'_{\ell_1 m}a'^{*}_{\ell_2 m}\rangle$, with $\ell_1\neq\ell_2$. Notably, the resulting leading corrections are of order $\beta \ell$ even in a full sky harmonic decomposition, contrary to what found instead for the $C_{\ell}$. This property of the off-diagonal terms of the covariance matrix has been recently used to forecast the ability of the Planck satellite to measure the magnitude and direction of our peculiar motion. In Ref. \cite{Amendola:2010ty}, for instance, it has been found perturbatively that a Planck-like experiment will be able to measure $\beta$ with an accuracy of about 30\%, and the direction of this motion with an error of about 20 degree. Ref.~\cite{Notari:2011sb}, instead, using a set of fitting formula for the $a'_{\ell m}$, shows that such an accuracy would be of about 20\% for Planck and even higher for a next generation of experiments.  

\subsection{Bispectrum}
In Eq.~(\ref{abkernel}) we assumed that the harmonic coefficients $a'_{\ell m}$ are measured by an observer moving along the direction ${\bf \hat{z}}$ coinciding with the $z$-axis of the harmonic decomposition (\ref{CMBrest}). Since the direction of our peculiar motion is identified with the direction of the CMB dipole, this choice corresponds to assume that the CMB dipole is aligned along the same $z$-axis. However, when including a mask in the analysis, {\it e.g.} to cut the background contaminated Galactic plane, it is particularly convenient  to work using Galactic coordinates and therefore identify the $z$-axis with the Galactic South Pole-North Pole direction. The latter direction does not experimentally coincide with the observed CMB dipole direction which is instead characterized by the longitude $l\simeq 249$ and the latitude $b\simeq 48$. This implies that the coefficients computed in Eq.~(\ref{boost2}) have to be rotated in order to consistently align the axis of the harmonic decomposition with the of Galactic South Pole-North Pole direction. In harmonic space this rotation is represented by a Wigner matrix $\mathcal{D} _{m m'}^{\ell}(\phi,\theta,\gamma)$, with properly chosen Euler angles $\phi$, $\theta$ and $\gamma$. After this Wigner rotation, the harmonic coefficients $a'_{\ell m}$ transform as follows
\beq
\hat{a}_{\ell m} = \sum_{-\ell \le m' \le \ell} \mathcal{D}_{m m'}^{\ell}(\phi,\theta,\gamma) \,a'_{\ell m'} 
\label{rot}
\eeq
where the coefficients $a'_{\ell m}$ were calculated in Eq.~(\ref{boost2}). The coefficients $\hat{a}_{\ell m}$ are now consistently expressed in the frame of an observer moving in the direction of the CMB dipole. 

These equations apply to a full sky harmonic decomposition and do not include yet any information about the window function $W({\bf \hat{n}})$ used to account for the partial sky coverage of current CMB experiments. 
If we denote by $\hat{T}({\bf \hat{n}})$ a full sky temperature map, its cut sky counterpart $\tilde{T}({\bf \hat{n}})$ takes the following form
\beqra
\tilde{T}({\bf \hat{n}}) &=& W({\bf \hat{n}}) \hat{T}({\bf \hat{n}}) \nonumber\\
&=& \sum_\ell \sum_{-\ell\le m\le\ell} w_{\ell m}Y_{\ell m}({\bf \hat{n}}) \hat{T}({\bf \hat{n}}) 
\eeqra
where we denoted by $w_{\ell m}$ the harmonic coefficients of the window function. Expanding also the full sky map 
\beq
\hat{T}(\hat{\n})= \sum_{lm} \hat{a}_{\ell m} Y_{lm}(\hat{\n})
\eeq 
and the cut sky map 
\beq
\tilde{T}(\hat{\n})= \sum_{\ell m} \tilde{a}_{\ell m} Y_{\ell m}(\hat{\n})
\eeq
one finds  
\beq
\tilde{a}_{\ell_{1}m_{1}} = \sum_{\ell_ {2}}\sum_{-\ell_{2}\le m_{2}\le\ell_{2}} \mathcal{F}_{\ell_{1}m_{1}\ell_{2}m_{2}} \,\hat{a}_{\ell_{2}m_{2}} 
\label{window}
\eeq
where the kernel $\mathcal{F}_{\ell_{1}m_{1}\ell_{2}m_{2}} $ relating the full sky coefficients $\hat{a}_{\ell_{2}m_{2}}$ to the cut sky coefficients $\tilde{a}_{\ell_{1}m_{1}}$ admits the following representation in terms of the Wigner 3-$j$ symbols
\beqra
 \mathcal{F}_{\ell_{1}m_{1}\ell_{2}m_{2}} &=& \sum_{\ell_ {3}}\sum_{-\ell_{3}\le m_{3}\le\ell_{3}} 
w_{\ell_{3} m_{3}} (-1)^{-m_{2}} \sqrt{\frac{(2\ell_{1}+1)(2\ell_{2}+1)(2\ell_{3}+1)}{4\pi}} \nonumber\\
\nonumber\\
&\times& 
 \left( \begin{array}{ccc}
 \ell_1 & \ell_2 & \ell_3 \\
 0 & 0 & 0 
 \end{array} \right)
 \left( \begin{array}{ccc}
 \ell_1 & \ell_2 & \ell_3 \\
 m_1 & -m_2 & m_3 
 \end{array} \right) \,.
\eeqra
The $\hat{a}_{\ell m}$ coefficients of Eq.~(\ref{window}) were computed in Eq.~(\ref{rot}). 
Using now the above equations, one can finally derive the three-point function $B^{m_1m_2m_3}_{\ell_1\,\ell_2\,\ell_3} \equiv \langle \tilde{a}_{\ell_1 m} \tilde{a}_{\ell_2 m} \tilde{a}_{\ell_3 m} \rangle$, namely the bispectrum, in the frame of an observer moving along the CMB dipole direction. In terms of the kernel $\mathcal{F}$ it reads as follows
\beq
B^{m_1m_2m_3}_{\ell_1\,\ell_2\,\ell_3} = \sum_{\ell m}\sum_{\ell' m'}\sum_{\ell'' m''} 
\mathcal{F}_{\ell_{1}m_{1}\ell m} \mathcal{F}_{\ell_{2}m_{2}\ell'm'} \mathcal{F}_{\ell_{3}m_{3}\ell''m''} 
\,\langle \hat{a}_{\ell m} \hat{a}_{\ell' m'}  \hat{a}_{\ell'' m''}\rangle \,.
\label{pseudo2}
\eeq
\section{Simulated temperature maps}
\label{data}
To quantitatively determine the significance of an aberration/Doppler induced Non-Gaussian signal, we need first a procedure to effectively simulate a large number of CMB maps observed in a reference frame moving along the direction of the CMB dipole. This procedure has been developed in Ref.~\cite{Catena:2012hq} and tested in the context of cosmological parameter estimation. In the following we will review this method and its implementation in a modified version of the HEALPix\footnote{http://healpix.jpl.nasa.gov} code, which we used to calculate the oscillatory angular integrals and the rotations in harmonic space introduced in the previous section. These simulated maps will be then analyzed in the next section, focusing on possible Non-Gaussian features. As a starting point our procedure requires a fiducial cosmological model for which the angular power spectrum $C^{(f)}_{\ell}$ is assumed to be known. The fiducial model assumed in this work is the WMAP 7-years best-fit model \cite{wmap7:params} 
Given a fiducial model, the map simulation follows five steps relying on the following HEALPix routines: 
\begin{itemize}
\item {\it synfast}: The synfast program generates full sky maps sampling from a fiducial model a realization of the $a_{\ell m}$ coefficients associated with the produced map. These are drawn from a Gaussian distribution with zero mean and variance equal to the $C^{(f)}_{\ell}$. We modified this code to allow the user to include the temperature transformation law given in Eq.~(\ref{boost}). We used this modified version of synfast to generate the temperature maps studied in the next section. (We restrict ourself to temperature maps, leaving for a future work an analysis devoted to aberration effects on CMB polarization maps).  
\item {\it anafast}: Temperature maps obtained with our modified version of synfast are then analyzed with the standard HEALPix version of the anafast code. This program expands these maps in spherical harmonics calculating the corresponding harmonic coefficients. These are the {\it exact} ({\it i.e.} non perturbative) $a'_{\ell m}$ coefficients of Eq.~(\ref{boost2}). 
\item {\it alteralm}: The $a'_{\ell m}$ coefficients are then rotated in a new reference frame, where the peculiar velocity inducing the aberration has the direction of the CMB dipole. To perform this rotation we use a modified version of the alteralm code which allows the user to choose the Euler angles characterizing the required rotation. The resulting coefficients are the $\hat{a}_{\ell m}$ given in Eq.~(\ref{rot}).  
\item {\it synfast}: A second call to the synfast program generates the temperature map $\hat{T}({\bf \hat{n}})$ associated with the $\hat{a}_{\ell m}$ coefficients.
\item {\it anafast:} Finally, we call again the anafast program to introduce the desired window function.
\end{itemize}
This procedure provides the CMB maps analyzed in the next section. 

\section{Analysis}

\label{analysis}

An interesting question related to aberration/Doppler effect is whether and how much it affects the estimation of primordial non-Gaussianity from inflation.
It is well known that different inflationary models can produce primordial bispectra with a model dependent shape, where the term ``shape'' is generally adopted to denote the functional dependence of a given bispectrum on different triangle configurations in Fourier/Harmonic space. So for example multi-field models of inflation can produce a``local'' shape type of bispectrum, where the NG signal in Fourier space is strongly peaked on $(k_1, k_2, k_3)$ triangles with $k_1 \ll k_2, \, k_3$ (the so called squeezed configurations). Single field models with non-standard kinetic terms in the inflaton Lagrangian can instead produce equilateral bispectra, where the signal
 peaks on configurations with $k_1 \sim k_2 \sim k_3$, and so on (see e.g. \cite{michele_review} and references therein for more information on bispectra from inflation).  For each shape, the strength of the primordial NG signal is usually parametrized by a single dimensionless parameter called $f_{\rm NL}$. A primordial bispectrum estimator measures the value of $f_{\rm NL}$ by fitting a given theoretical shape to the 3-point function extracted from the data. 
In the idealized case of full sky CMB measurements and homogeneous noise, an optimal bispectrum estimator of $f_{\rm NL}$ is derived as (see e.g. \cite{michele_review}) .
\begin{eqnarray}
	\hat{f}_{\rm NL}&=&\frac{1}{\mathcal{N}} \sum_{\{l_i ,m_i\}} 	
		\frac{\mathcal{G}_{l_1,l_2,l_3}^{m_1,m_2,m_3}b_{l_1,l_2,l_3}^{f_{\rm NL}=1}}{C_{l_1} C_{l_2} C_{l_3}}
		a_{l_1,m_1}a_{l_2,m_2}a_{l_3,m_3}, \nonumber\\
	\mathcal{N}&=&\sum_{\{l_i ,m_i\}}
		\frac{\left(\mathcal{G}_{l_1,l_2,l_3}^{m_1,m_2,m_3}b_{l_1,l_2,l_3}^{f_{\rm NL}=1}\right)^2}{C_{l_1} C_{l_2} C_{l_3}}\label{eqn:optimalbisp1},
\end{eqnarray}
where $b_{l_1,l_2,l_3}$ is the theoretical reduced bispectrum for a given inflationary model (obtained through a convolution of the 
primordial shape with CMB radiation transfer functions) and $\mathcal{G}_{l_1,l_2,l_3}^{m_1,m_2,m_3}$ is the Gaunt integral.
A brute force implementation of formula (\ref{eqn:optimalbisp1}) is numerically unfeasible for modern satellite CMB experiments (like WMAP or Planck) due to the huge number of configurations to evaluate.  Komatsu, Spergel and Wandelt showed in \cite{komatsu.et.al.2005} that when the primordial bispectrum template $b_{l_1,l_2,l_3}$ is separable (i.e. it is a linear combination of products of functions of only one of the three $l$) then equation (\ref{eqn:optimalbisp1}) can be conveniently factorized and computed in a much smaller number of operations than a brute force evaluatin would require. This allow to construct an $f_{\rm NL}$ estimator (from now on KSW) that can be applied also to large CMB datasets.
Suitable separable approximations have been found for the 3 most important shapes in the literature: the local, equilateral and orthogonal shape \cite{Creminellietal}.  Following the arguments in e.g. \cite{yadav2008} we can then write the KSW estimator for these three shapes as:
\begin{equation}
  f_{\rm NL}^{\mathrm{shape}} = S_{\mathrm{shape}} / F_{\mathrm{shape}} \; ,
\end{equation}
where $F_{\mathrm{shape}}$ is the diagonal of the Fisher matrix for the three shapes \cite{komatsu.et.al.2005}, while $S$ is defined as follow:
\begin{eqnarray}
  S_{\rm local} &=& \int r^2 dr \int d^2 \hat{\bf n}\left[A(\hat{\bf n},r)B^2(\hat{\bf n},r)-2B(\hat{\bf n},r)\left\langle A(\hat{\bf n},r)B(\hat{\bf n},r)\right\rangle_{MC}\right. \nonumber\\
  &&\left. -A(\hat{\bf n},r)\left\langle B^2(\hat{\bf n},r)\right\rangle_{MC}\right], \label{s_local}\\
  S_{\rm equi} &=& -3 S_{\rm local} +6\int r^2 dr \int d^2 \hat{\bf n}
  \left\lbrace B(\hat{\bf n},r)C(\hat{\bf n},r)D(\hat{\bf n},r)-B(\hat{\bf n},r)\left\langle C(\hat{\bf n},r)D(\hat{\bf n},r)\right\rangle_{MC}\right. \nonumber\\
  &&\left. -C(\hat{\bf n},r)\left\langle B(\hat{\bf n},r)D(\hat{\bf n},r)\right\rangle_{MC} - D(\hat{\bf n},r)\left\langle B(\hat{\bf n},r)C(\hat{\bf n},r)\right\rangle_{MC} \right. \nonumber\\
  &&\left. -\frac{1}{3}\left[D^{3}(\hat{\bf n},r) -3D(\hat{\bf n},r)\left\langle D^{2}(\hat{\bf n},r)\right\rangle_{MC} \right] \right\rbrace, \label{s_equi}\\
  S_{\rm ortho} &=& 3S_{\rm equi} -2 \int r^2 dr \int d^2 \hat{\bf n} \left[D^{3}(\hat{\bf n},r) -3D(\hat{\bf n},r)\left\langle D^{2}(\hat{\bf n},r)\right\rangle_{MC} \right] \label{s_ortho} \; ,
\end{eqnarray}
and the filtered maps $A,B,C,D$ are defined as:
\begin{eqnarray}
  A(\hat{\mathbf{n}},r) &\equiv & \sum_{lm} b_l \alpha_l (r) (C^{-1} a)_{lm}Y_{lm}(\hat{\mathbf{n}}),\label{a_matrix}\\
  B(\hat{\mathbf{n}},r) &\equiv &\sum_{lm} b_l \beta_l (r) (C^{-1} a)_{lm}Y_{lm}(\hat{\mathbf{n}}),\label{b_matrix}\\
  C(\hat{\mathbf{n}},r) &\equiv & \sum_{lm} b_l \gamma_l (r) (C^{-1} a)_{lm}Y_{lm}(\hat{\mathbf{n}}),\label{c_matrix}\\
  D(\hat{\mathbf{n}},r) &\equiv &\sum_{lm} b_l \delta_l (r) (C^{-1} a)_{lm}Y_{lm}(\hat{\mathbf{n}}),\label{d_matrix}
\end{eqnarray}
In the formulae above, $b_l$ is the experimental beam, $C$ is the total power spectrum including the CMB signal, $C_{l}^{CMB}$, and experimental noise, $N_l$, ($C_{l} \equiv C_{l}^{CMB} b_{l}^2+ N_l$); finally we defined $\alpha, \beta, \gamma, \delta$ as:
\begin{eqnarray}
  \alpha_l (r) &=& \frac{2}{\pi}\int k^2 dk g_{Tl}(k)j_l (kr),\label{alfa}\\
  \beta_l (r) &=& \frac{2}{\pi}\int k^2 dk P_{\Phi}(k)g_{Tl}(k)j_l (kr),\label{beta}\\
  \gamma_l (r) &=& \frac{2}{\pi}\int k^2 dk P_{\Phi}^{1/3}(k)g_{Tl}(k)j_l (kr),\label{gamma}\\
  \delta_l (r) &=& \frac{2}{\pi}\int k^2 dk P_{\Phi}^{2/3}(k)g_{Tl}(k)j_l (kr)\label{delta} \; ,
\end{eqnarray}
$g_{Tl}$ is the temperature radiation transfer function\footnote{Obtained from a modified version of CAMB code, http://camb.info} and $P_{\Phi}$ is the primordial curvature perturbation power spectrum.
In the formula \ref{s_local}, \ref{s_equi} and \ref{s_ortho}, MC denotes Monte-Carlo averages over CMB simulations including all experimental features like noise or beam. Those MC averages appear in terms that are linear in the $a_{lm}$'s so that they are zero on average. Such terms do not appear in the idealized cubic estimator of formula (\ref{eqn:optimalbisp1}). The reason is that the formula above defines an optimal $f_{\rm NL}$ estimator only in the case of statistically isotropic data. As originally shown in \cite{Creminellietal}, the linear terms have to be included in order to make the estimator optimal in the realistic case when statistically anisotropic effects are present in the data, in particular sky cut and anisotropic noise. 

In the following we will study the effect of CMB aberration/Doppler on $f_{\rm NL}$ by building a KSW estimator and applying it on simulations including aberration/Doppler. The only difference between our implementation and the optimal KSW formulae above is that we will replace the inverse covariance weighting of the data $C^{-1} a$ with a simple diagonal approximation ${a_{lm}/C_l}$. This will make the estimator only slightly suboptimal when a sky cut is applied, while strongly simplifying the numerical implementation.

\subsection{\label{ksw_sect}Methodology}
In order to assess the effect of CMB aberration on primordial NG we estimate local, equilateral and orthogonal $f_{\rm NL}$ from 120 Gaussian maps without aberration/Doppler and compare to the same Gaussian maps where the aberration/Doppler is included. 
This is done in the case of full sky and with a mask (KQ75 WMAP7 mask defined in \cite{gold_wmap7}) for an ideal experiment (no beam and no noise).

\subsection{\label{fullsky}Full sky results}
\begin{figure} [t]
\begin{center}
\includegraphics[width=130mm,height=60mm]{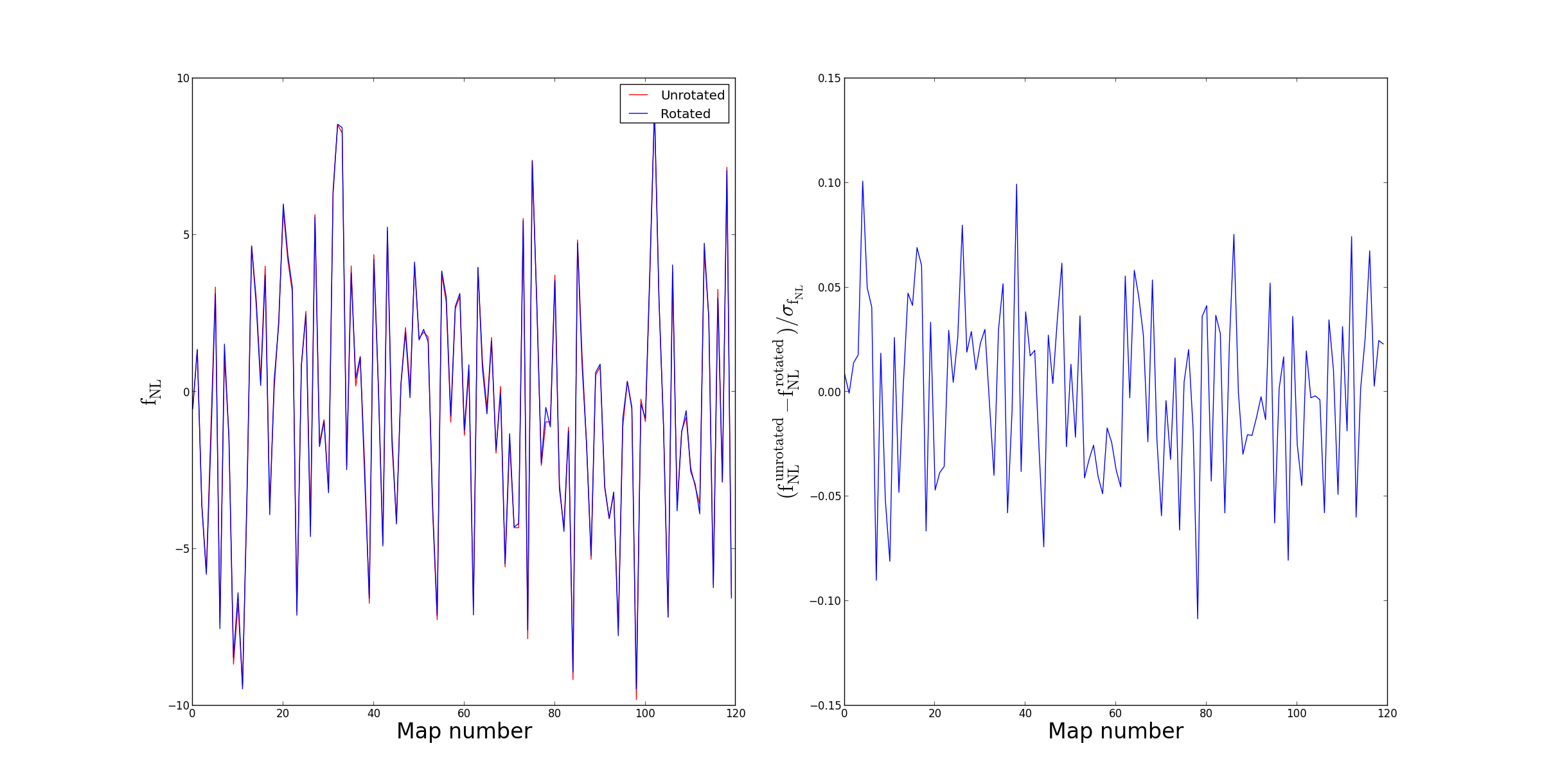}
\end{center}
\caption{Full sky, local configuration.}
\label{ideal_local}
\end{figure}

\begin{figure} [t]
\begin{center}
\includegraphics[width=130mm,height=60mm]{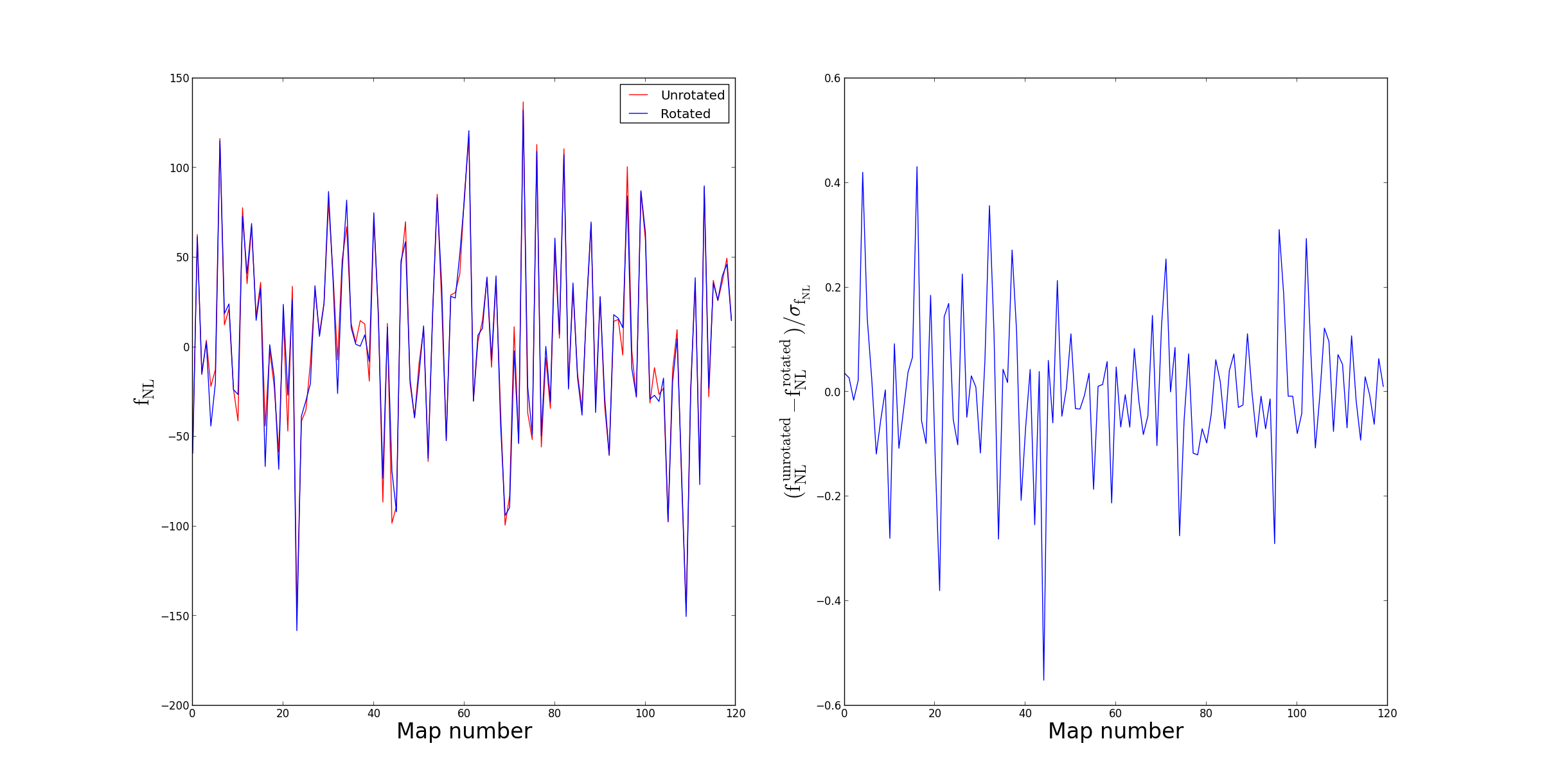}
\end{center}
\caption{Full sky, equilateral configuration.}
\label{ideal_equi}
\end{figure}

\begin{figure} [t]
\begin{center}
\includegraphics[width=130mm,height=60mm]{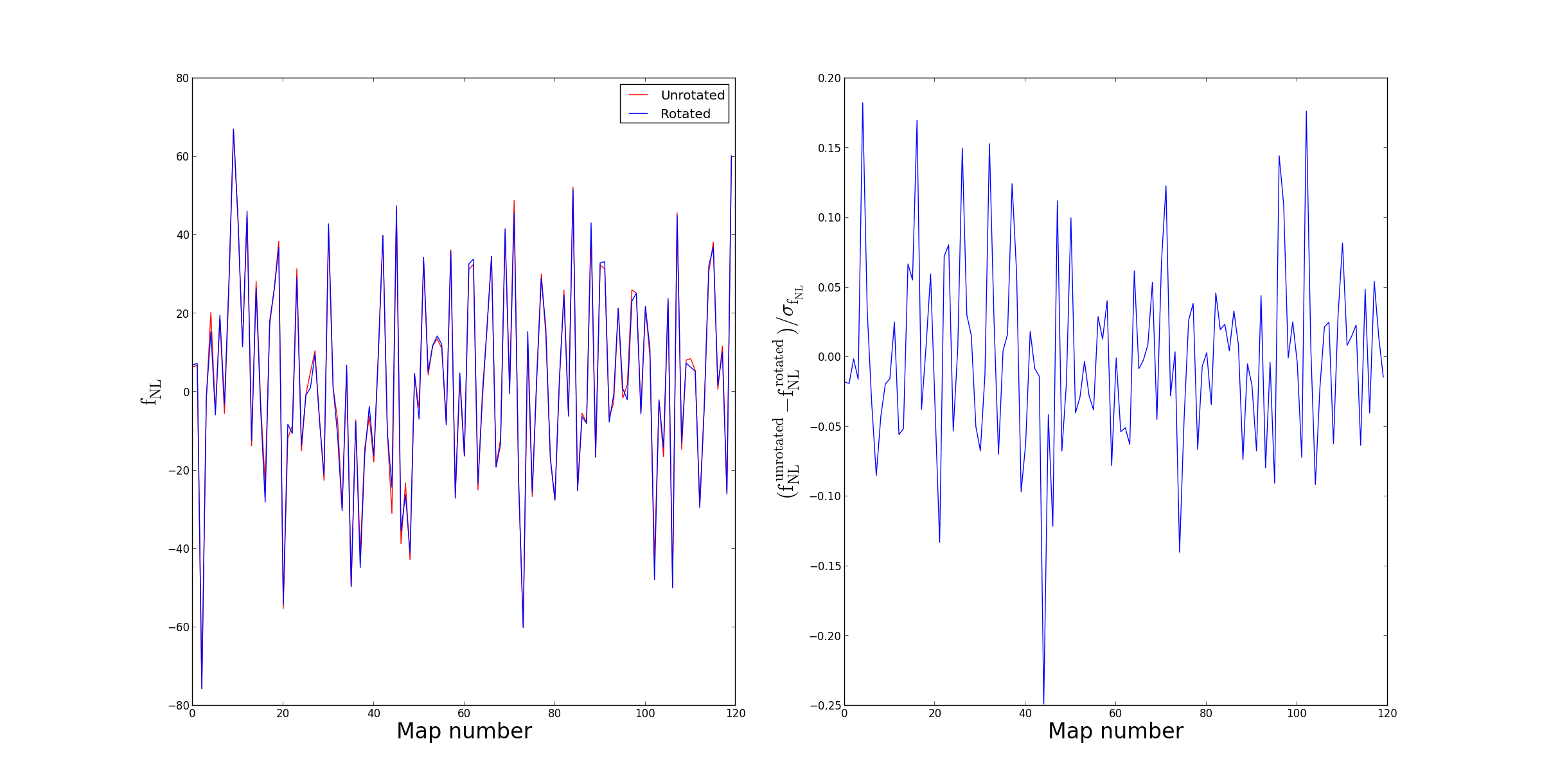}
\end{center}
\caption{Full sky, orthogonal configuration.}
\label{ideal_ortho}
\end{figure}

Results are shown in figures \ref{ideal_local},\ref{ideal_equi} and \ref{ideal_ortho}. In the left panel we plot the value of $f_{\rm NL}$ for all the shapes and for all the maps. Since the maps are Gaussian the $f_{\rm NL}$ values must be consistent with zero within the standard deviations: $\sigma_{f_{\rm NL}^{\rm local}}=4.2$, $\sigma_{f_{\rm NL}^{\rm equilateral}}=52.3$, $\sigma_{f_{\rm NL}^{\rm orthogonal}}=26.3$. That of course holds for both sets of maps (with or without the aberration and Doppler effects), as they are both Gaussian. However correlations induced by CMB aberration and Doppler can change the estimator variance, and the measured $f_{NL}$ values on a map-by-map basis. In blue we plot the $f_{\rm NL}$ values from the Gaussian maps without aberration and Doppler effects, and in red the values from the same simulations with aberration and Doppler included. The overall effect is small. In order to show the small deviations more clearly we plot in the right panels the difference between the $f_{\rm NL}$ obtained from a Gaussian map and the $f_{\rm NL}$ from the same map with aberration and Doppler in terms of standard deviations. For all the cases the difference is less than half sigma. 

We can conclude that for the case of no mask we didn't find any deviation from the case of no aberration and Doppler. 


\subsection{Cut sky results}
\begin{figure}[t]
\begin{center}
\includegraphics[width=130mm,height=60mm]{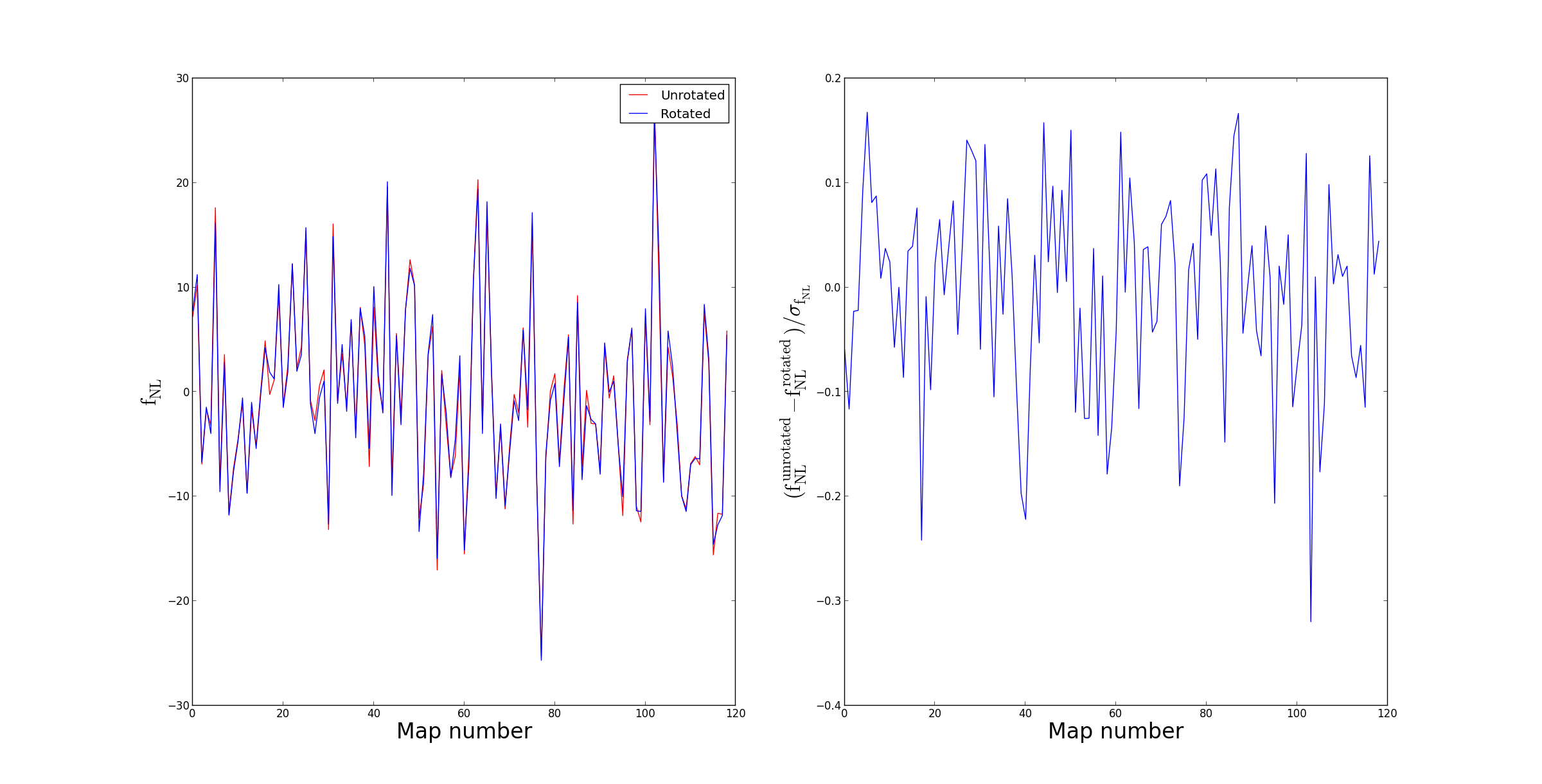}
\end{center}
\caption{Cut sky, local configuration.}
\label{ideal_m_local}
\end{figure}

\begin{figure}[t]
\begin{center}
\includegraphics[width=130mm,height=60mm]{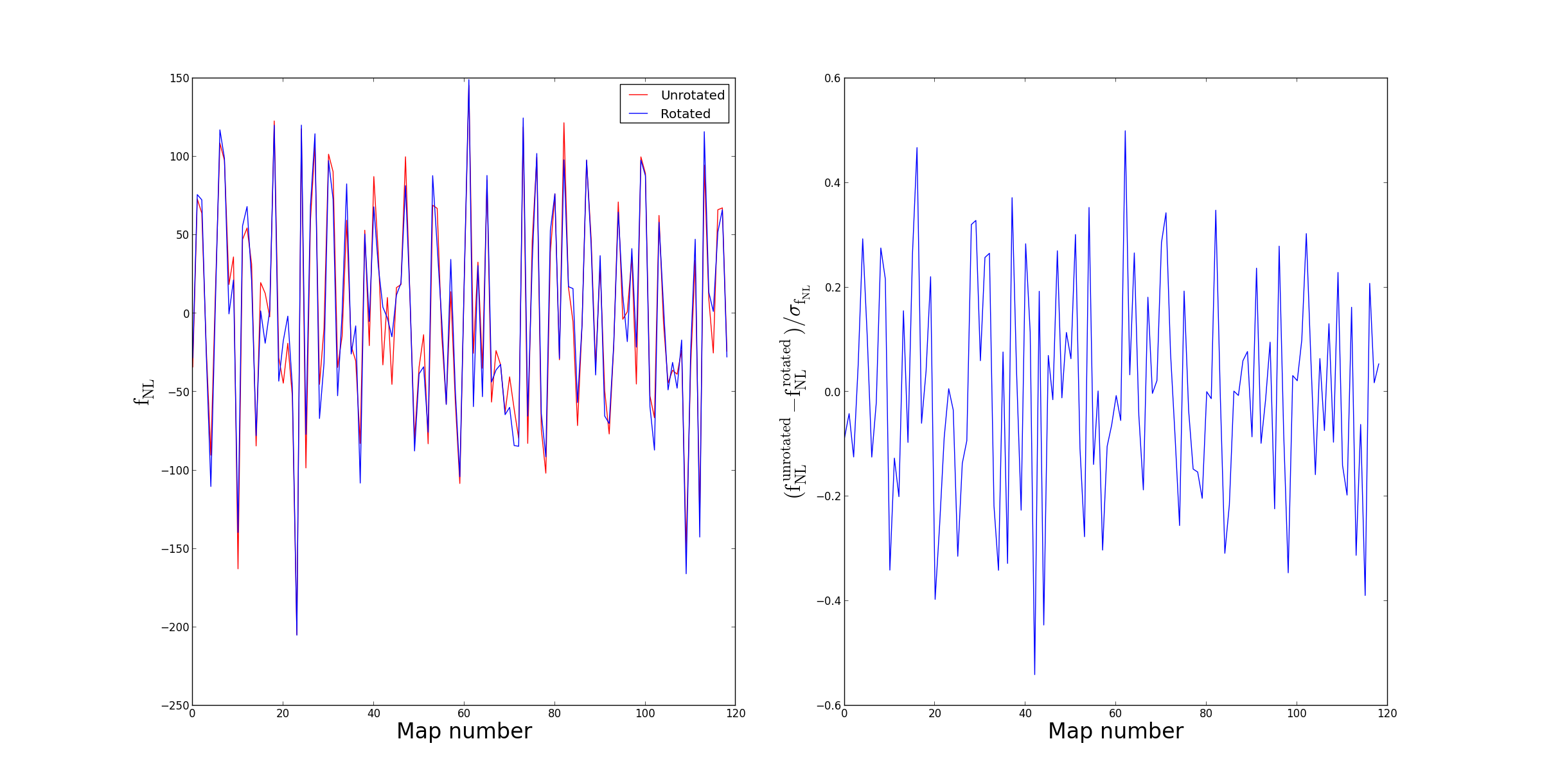}
\end{center}
\caption{Cut sky equilateral configuration.}
\label{ideal_m_equi}
\end{figure}

\begin{figure}[t]
\begin{center}
\includegraphics[width=130mm,height=60mm]{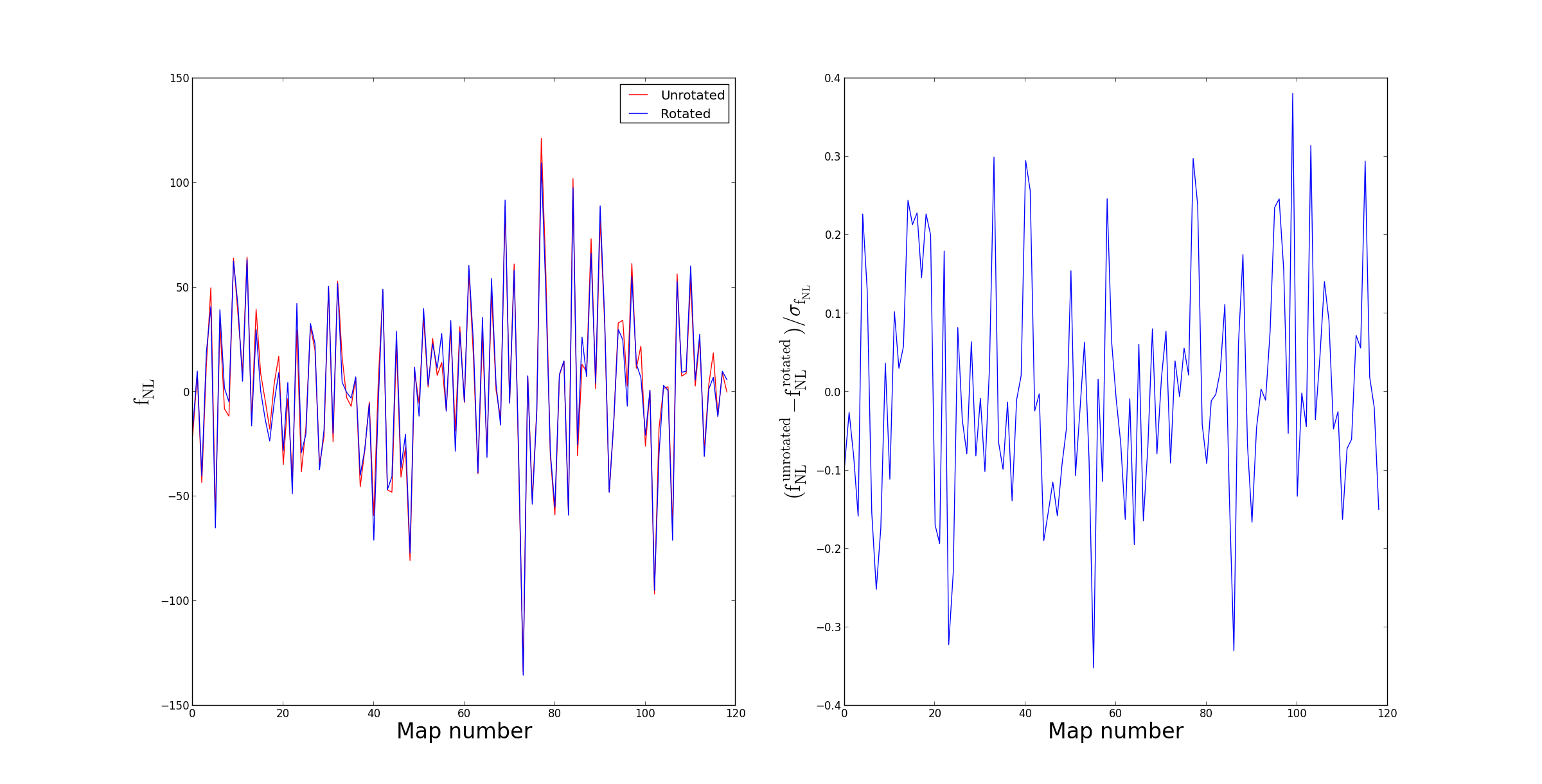}
\end{center}
\caption{Cut sky, orthogonal configuration.}
\label{ideal_m_ortho}
\end{figure}

The same tests performed in section \ref{fullsky} are done in this section but with the KQ75 mask applied, in order to verify if the presence of a sky cut can enhance the differences. The starting Gaussian maps are the same as in the previous section. The presence of a sky cut increases the standard deviations to: $\sigma_{f_{\rm NL}^{\rm local}}=8.7$, $\sigma_{f_{\rm NL}^{\rm equilateral}}=67.5$, $\sigma_{f_{\rm NL}^{\rm orthogonal}}=39.8$. Results are shown in figure \ref{ideal_m_local},\ref{ideal_m_equi} and \ref{ideal_m_ortho}. Even in the cut sky case there is no evidence of a significant impact of aberration and Doppler on $f_{\rm NL}$.




\section{Discussion and conclusions}
\label{conclusion}
In the present paper we have studied the impact of the motion of a peculiar observer, with a velocity $\beta\equiv v/c= 1.23 \times 10^{-3}$, on Gaussian maps of the CMB sky with Planck-like resolution (lmax $= 2000$) in order to see if this could significantly bias the measured value of the Non-Gaussianity parameter $f_{NL}$.
In fact a peculiar motion induces aberration and Doppler effects on the CMB maps, which mixes multipoles with coefficients of order $\beta \ell$, which are therefore ${\cal O}(1)$ at $\ell \gtrsim 800$ and so the measured $a_{\ell m}$ are actually very different from the rest-frame ones. Such mixing is still a linear transformation, so on average its effect on 3-point function vanishes, but there could be {\it a priori} a sizable shift on a single random realization of a map.
We simulated then 120 maps with Planck-like resolution and we have checked for each map how much $f_{NL}$ can change by applying a boost transformation, using estimators for local, equilateral and orthogonal Non-Gaussianity.

We find that the shift in $f_{NL}$ is always at most about half standard deviation, both for maps with full-sky coverage and with a mask.
So, our conclusion  is that we can safely analyze the map in the boosted frame {\it if } our original map is Gaussian, without significantly biasing the $f_{NL}$ measurement.
However more work is needed to assess the case in which the map is already Non-Gaussian in the CMB rest frame: in such a case there could be a considerable shift due to the boost transformation, because the  mixing matrix to be applied because of aberration and Doppler effects would act on non-zero 3-point functions.

\acknowledgments
Some of the results in this paper have been derived using the HEALPix \cite{Gorski:2004by} package.

\end{document}